\documentclass[prb,amsmath,amssymb,superscriptaddress]{revtex4}
\usepackage{graphicx}
\usepackage{dcolumn}
\usepackage{color}

\begin{document}


\title{Ferromagnetic resonance in $\epsilon$-Co magnetic composites}



\author{Khattiya Chalapat}
\affiliation{O. V. Lounasmaa Laboratory, Aalto University, P.O. Box 15100, FI-00076, Finland}
\author{Jaakko V. I. Timonen}
\affiliation{Department of Applied Physics, Aalto University, P.O. Box 14100, FI-00076, Finland}
\author{Maija Huuppola}
\affiliation{Department of Chemistry, Aalto University, P.O. Box 16100, FI-00076, Finland}
\author{Lari Koponen}
\affiliation{Department of Applied Physics, Aalto University, P.O. Box 14100, FI-00076, Finland}
\author{Christoffer Johans}
\affiliation{Department of Chemistry, Aalto University, P.O. Box 16100, FI-00076, Finland}
\author{Robin H. A. Ras}
\affiliation{Department of Applied Physics, Aalto University, P.O. Box 14100, FI-00076, Finland}
\author{Olli Ikkala}
\affiliation{Department of Applied Physics, Aalto University, P.O. Box 14100, FI-00076, Finland}
\author{Markku A. Oksanen}
\affiliation{Nokia Research Center, P.O. Box 407, 00045 NOKIA GROUP, Finland}
\author{Eira Sepp\"al\"a}\altaffiliation{Current affiliation: Spinverse Oy, Tekniikantie 14, 02150 Espoo, Finland}
\affiliation{Nokia Research Center, P.O. Box 407, 00045 NOKIA GROUP, Finland}
\author{G. S. Paraoanu}
\affiliation{O. V. Lounasmaa Laboratory, Aalto University, P.O. Box 15100, FI-00076, Finland}


\begin{abstract}

We investigate the electromagnetic properties of assemblies of nanoscale $\epsilon$-cobalt crystals with size range between 5 nm to 35 nm, embedded in a polystyrene (PS) matrix, at microwave (1-12 GHz) frequencies. We investigate the samples by transmission electron microscopy (TEM) imaging, demonstrating that the particles aggregate and form chains and clusters. By using a broadband coaxial-line method, we extract the magnetic permeability in the frequency range from 1 to 12 GHz, and we study the shift of the ferromagnetic resonance  with respect to an externally applied magnetic field.  We find that the zero-magnetic field ferromagnetic resonant peak shifts towards higher frequencies at finite magnetic fields, and the magnitude of complex permeability is reduced. At fields larger than 2.5 kOe the resonant frequency changes linearly with the applied magnetic field, demonstrating the transition to a state in which the nanoparticles become dynamically decoupled. In this regime, the particles inside clusters can be treated as non-interacting, and the peak position can be predicted from  Kittel's ferromagnetic resonance theory for non-interacting uniaxial spherical particles combined with the Landau-Lifshitz-Gilbert (LLG) equation. In contrast, at low magnetic fields this magnetic order breaks down and the resonant frequency in zero magnetic field reaches a saturation value reflecting the interparticle interactions as resulting from aggregation. Our results show that the electromagnetic properties of these composite materials can be tuned by external magnetic fields and by changes in the aggregation structure.

\end{abstract}


\maketitle

\section{Introduction}

The realization of novel materials with designed and tunable optical, mechanical, thermal and electromagnetic properties is a major goal in nanotechnology. A promising direction is the study of
nanocomposites, consisting of nano-scale particles (silica, Fe, Co, Ni, Al, Zn, Ti, as well as their oxides) or other low-dimensional structures (carbon nanotubes, graphene, DNA) embedded in a bulk matrix or polymer \cite{Ajayan}. Nanocomposites with magnetic properties, which use magnetic nanoparticles inserted in dielectric matrices, are of considerable technological importance, due to the simplicity of fabrication and potential applications for radio-frequency and microwave antennas, for GHz-frequency data transfer components, and for electromagnetic shielding \cite{Chen2004,Timonen2010}.
On the scientific side, these materials are intriguing because concepts such as ferromagnetic resonance (FMR), predicted almost a century ago \cite{LandauLifshitz,Kittel1947,Kittel1948} and later demonstrated experimentally in bulk materials \cite{Griffiths1946, Yager1947}, can now be applied to particles with nano-scale dimensions and are relevant for understanding the properties of the resulting composite materials. The standard measurement technique in FMR is narrow-band: the material is placed in a microwave cavity whose response is monitored while sweeping the external magnetizing field until the lowest quality factor (maximum of absorption) is observed. However, for modern applications a broadband characterization of the electromagnetic properties of the material is required. Recently, non-resonant transmission line methods have been employed to study magnetic nanoparticles, but, with the notable exception of magnetic fluids in the 1990's \cite{Fannin1996,Fannin1999}, only in zero external magnetic field \cite{Wu2006,ZhengHong2009,YangYong2010,Neo2010,LonggangYan2010,Yang2011}. There is to date no systematic characterization of the nano-scale and magnetic properties (including the effects of sizes, shapes, magnetic domains, aggregation, {\it etc.}) of composite materials containing magnetic nanoparticles.

In this paper we present a comprehensive study of the broadband microwave properties and FMR resonance of composite materials comprising chemically synthesized $\epsilon$-cobalt nanoparticles with
sizes between 5 nm and 35 nm embedded in polystyrene, both in zero and in a finite magnetic field, including information from superconducting quantum interference device (SQUID) measurements. The particles are synthesized by two different methods, the hot-injection method \cite{Puntes} and the heating-up method \cite{Timonen2011}. At macroscopic scales, bulk cobalt has only two forms of lattice structures, namely hexagonal-closed-packed (hcp) and face-centered-cubic (fcc). For nanoparticles, a cubic $\epsilon$-cobalt phase, with a structure similar to $\beta$-manganese, has been observed in addition to the hcp and fcc phases \cite{Dinega1999, Sun1999}, and more recently developed chemical syntheses have allowed the production of $\epsilon$-cobalt particles at specific diameters (narrow size distribution) \cite{Puntes2001, Xia2009}. The resonance condition of the cobalt nanocrystals is found by combining Kittel's FMR theory \cite{Kittel1947,Kittel1948} with the Landau-Lifshitz-Gilbert (LLG) equation \cite{LandauLifshitz,Gilbert2004}
and various effective-medium models. Kittel's theory is a valid approach under the
assumption that the eddy current skin depths are larger than the typical dimension of the particles. In Co, the skin depths in the
frequency range between 1 GHz and 10 GHz are of the order of 100 nm, therefore the nanoparticles used in our samples satisfy this
requirement.  We identify two distinct magnetization regimes: at high external magnetic fields, the FMR resonance changes linearly with the field and the composite is well described by the Kittel-LLG model, while at low magnetic field the FMR resonance saturates to a constant value. This is explained by the existence of
particle-particle magnetic interactions, which become the dominant effect at small and zero external magnetic field.
Transmission electron microscopy (TEM) imaging confirms the formation of particle aggregates in the composite.


\section{Methods}
\label{methods}

\subsection{Synthesis and TEM imaging of $\epsilon$-cobalt nanoparticles}
Spherical $\epsilon$-cobalt nanoparticles were synthesized by thermally decomposing 1080 mg of dicobalt octacarbonyl (STREM, 95\%) in presence of 200 mg of trioctylphosphine oxide (Sigma-Aldrich, 99\%) and 360 mg of oleic acid (Sigma-Aldrich, 99\%) in 30 ml of 1,2-dichlorobenzene (Sigma-Aldrich, 99\%) under inert nitrogen atmosphere \cite{Puntes2001, Timonen2011}. The cobalt precursor was either injected into the surfactant mixture at 180 $^o$C \cite{Puntes} or mixed with the surfactant solution at room temperature and heated up to 180 $^o$C \cite{Timonen2011}. No difference in particles produced by the hot-injection or heating-up methods was observed. The particle size was adjusted by the temperature kinetics during the reaction \cite{Timonen2011}, {\it i.e.} by using a higher heating rate in the heating-up method or a higher recovery rate in the hot-injection method to increase the nucleation rate and to produce smaller particles. The excess surfactants were removed after synthesis by precipitating the particles by adding methanol, followed by redispersing the particles into toluene. The composites were fabricated by dissolving desired amounts of polystyrene (Sigma-Aldrich) in the nanoparticle dispersion, followed by evaporation of the solvent to yield black solid composite materials. The samples can be regarded as a two-component
composite, made of magnetically-active $\rm Co$ particles (density $\rho_{\rm Co} = 8.63$ g/cm$^3$) in a magnetically-inert polystyrene bulk (density $\rho_{\rm PS} = 1.05$ g/cm$^3$). The amount of oleic acid left is small and its density is anyway close to that of polystyrene ($\rho_{\rm oleic~acid} = 0.89$ g/cm$^3$). Both the pure nanoparticles and thin cross-sections of the composite materials were analyzed by transmission electron microscopy (Tecnai T12 and JEOL JEM-3200FSC). The images showed that our $\epsilon$-cobalt nanoparticles typically consists of a single crystal core. For the high-frequency measurements, the composite materials were compression molded at 150$^o$ C into circular pellets with a diameter of 7.0 mm and thickness of 4.0 mm. A circular hole of 3.0 mm was drilled through each pellet to accommodate the inner conductor of the coaxial line.

\subsection{Theoretical model}

The phenomenon of ferromagnetic resonance (FMR) was predicted by Landau and Lifshitz in 1935 \cite{LandauLifshitz} and was observed experimentally more than a decade after by Griffiths \cite{Griffiths1946}, and then by W. A. Yager and R. M. Bozorth  \cite{Yager1947}.
Griffith also found that the ferromagnetic resonance does not occur exactly at the electron spin resonance of frequency
$\hbar\omega_{0} = g_{e}\mu_B H$ or $\omega_{0} = \gamma H$, where $g_{e}$ is the electron g-factor,
$\mu_B = e\hbar/2m_e$ is the Bohr magneton, $H$ is the internal magnetizing field, and $\gamma_{e} = g_{e} e/2 m_e$ is the
electron gyromagnetic ratio, $\gamma_{e}/2\pi$ = 2.7992$\times10^{10}$ Hz/T. Immediately after, Kittel \cite{Kittel1947,Kittel1948} proposed that the ferromagnetic resonance condition should be modified
from the original Landau-Lifshitz theory by taking into account the shape and crystalline anisotropy through the demagnetizing fields.

In the case  of disordered magnetic composites, the physics is expected to be very complicated, due the interplay of geometry, anisotropy, interparticle interactions, and sample inhomogeneities. In principle, one can attempt to give a microscopic description of these effects and average over several realizations; however this approach is likely to be complicated. Here we propose a simple phenomenological description of the composites. Similar to effective-field theories in physics, the idea is to construct a model that incorporates the basic known physical processes - in this case, magnetization precession in an applied magnetic field and dissipation - and solve the problem using very general methods - linear response theory in our case. This allows us on one hand to extract the electromagnetic parameters that are relevant for technological applications (high-frequency magnetic permittivity and permeability), and on the other hand to give an effective quantitative  description of the ferromagnetic resonance observed.

The starting point of the model is the observation that if the external magnetic field $H$ applied is large, then
the particles will get strongly magnetized in the direction of the applied field. As a result, irrespective to the configuration of the sample or on the geometry of particle anisotropy, the sample behaves as a collection of noninteracting domains with magnetization rotating in synchronization around a total effective magnetic field $H_{\rm eff}= H + \tilde{H}$. In this regime, one then expects a linear law for the
resonance  frequency,
\begin{equation}
\omega_0 = \gamma H_{\rm eff} = \gamma (H + \tilde{H}), \label{linear}
\end{equation}
where $\tilde{H}$ incorporates the effects mentioned above. As we will see later, the main contribution in $\tilde{H}$ comes from an effective magnetic anisotropy field $H_{\rm A}$.

The rotation of the magnetic domains is also accompanied by energy loss, which can be accounted for through the Landau, Lifshitz \cite{LandauLifshitz} and Gilbert \cite{Gilbert2004} formalism. The dynamics of magnetization is described by the Landau-Lifshitz-Gilbert (LLG) equation
\begin{equation}
\frac{d\vec{M}_{\rm tot}}{dt} = - \gamma \vec{M}_{\rm tot} \times \vec{H}_{\rm tot} +\frac{\alpha}{M_s}\vec{M}_{\rm tot} \times \frac{d\vec{M}_{\rm tot}}{dt} \label{LLG}.
\end{equation}
The first term on the right hand side represents the precession of magnetization. The energy loss is taken into account by the second term via a single damping constant $\alpha$. $\vec{M}_{\rm tot}$ is the magnetization of the system, and $\vec{H}_{\rm tot}$ is the total magnetizing field which includes a small microwave probe field ($h e^{i\omega t}$),
\begin{equation}
\vec{H}_{\rm tot} = \vec{H}_{\rm eff} + \vec{h} e^{i\omega t}.\label{htot}
\end{equation}
Demagnetizing fields can be introduced as well in Eq. (\ref{htot}) and it can be checked that for spherical symmetry they cancel out from the final results. Similarly, the magnetization at a given time includes the magnetization of the particle $\approx \vec{M_s}$ and the time-varying term,
\begin{equation}
\vec{M}_{\rm tot} = \vec{M}_s+ \vec{m} e^{i\omega t},
\end{equation}
where $\vec{M_s}$ is the saturation magnetization.

To solve the equation, it is convenient to take the coordinate $z$ along the direction of the field $\vec{H}_{\rm eff}$, which in the limit of large fields discussed above coincides with the direction of the magnetization $M_{s}$.  The magnetic susceptibility is given by $\chi_{\perp} = \partial m_x/\partial h_x = \partial m_y/\partial h_y$, with the magnitude $h$ of the probe microwave field assumed much smaller than the static fields, $h\ll H_{\rm eff}$. This yields a simple analytical expression for the complex susceptibility $\chi_{\perp}=\chi_{\perp}'-i\chi_{\perp} ''$,
\begin{equation}
\chi_{\perp}' = \frac{\gamma M_s \omega_0 [ \omega_0^2 - \omega^2(1-\alpha^2)]}{[\omega_0^2 - \omega^2 (1 + \alpha^2)]^2 + 4\omega^2\omega_0^2\alpha^2},
\label{LLGReMuParticle}
\end{equation}
\begin{equation}
\chi_{\perp} '' = \frac{\alpha \gamma M_s \omega [ \omega_0^2 + \omega^2(1+\alpha^2)]}{[\omega_0^2 - \omega^2 (1 + \alpha^2)]^2 + 4\omega^2\omega_0^2\alpha^2}, \label{LLGImMuParticle}
\end{equation}
where $\omega_0 = \gamma(H+\tilde{H})$.

The measured resonance frequency in the presence of dissipation can be obtained by searching for the maximum of Eq. (\ref{LLGImMuParticle}). We have solved the equation $\partial \chi_{\perp} ''/\partial \omega =0$ analytically, using Mathematica. The polynomial resulting from the derivative has 6 roots, of which we select the physically relevant one (real and positive), and check that it corresponds to a maximum. This solution is
\begin{equation}
\omega_{r} = 2\pi f_{r} = \frac{\omega_{0}}{\sqrt{1 + \alpha^2}} = \frac{\gamma}{\sqrt{1 + \alpha^2}}(H+\tilde{H}).
\label{rezz}
\end{equation}
This solution is also intuitively appealing, as it corresponds to a zero in the first term in the denominator of Eq. (\ref{LLGImMuParticle}), a term which tends to increase fast if the frequency is detuned only slightly from the resonance.

In general, the total susceptibility is $\chi(\omega) = \chi(\omega)' - i\chi(\omega)''$ and the relative permeability $\mu_{\rm Co}(\omega) = 1 + \chi (\omega )$ is composed of both parallel $\chi_\parallel$, and perpendicular $\chi_\perp$ components. The total susceptibility is an average of these components, with
\begin{equation}
\chi(\omega) = \frac{1}{3}[\chi_\parallel(\omega) + 2 \chi_\perp(\omega)].
\end{equation}
The perpendicular component can be well described by the LLG-Kittel theory, {\it i.e.}  Eq. (\ref{LLGReMuParticle})-(\ref{LLGImMuParticle}). The parallel component is purely relaxational and usually assumed to be described by the Debye model \cite{Fannin1997},
\begin{equation}
\chi_\parallel(\omega) = \frac{\chi_\parallel(0)}{1+i \omega\tau_\parallel}.
\label{DebyeMu}
\end{equation}
Here $\tau_\parallel = \tau_0 \sigma$, where
$\tau_0 = (\gamma \alpha_{\parallel} H_{\rm A} )^{-1}$, $\sigma = K V/ k_B T$, $V$ is the average particle volume, and the effective anisotropy field $H_{\rm A}$ and the effective anisotropy constant $K$ can be estimated from SQUID measurements (see subsection II C).

For materials with inclusions, mixing rules such as Bruggeman equation \cite{Bruggeman1935} and the Maxwell-Garnett model \cite{Garnett1904,Mallet2005} give good results when the volume fraction of the inclusions is not too large. In the case of spherical inclusions, the Maxwell-Garnett formalism gives the following expression for permeability
\begin{equation}
\mu_{r} = \frac{2(f-1)\mu_{\rm M}^2 - (1+2f) \mu_{\rm Co} \mu_{\rm M} }{ (f-1)
\mu_{\rm Co} - (2+f)\mu_{\rm M}},
\label{MuCompositeMaxwell}
\end{equation}
while the Bruggeman equation reads
\begin{eqnarray}
\mu_{r} &= &\frac{1}{4}[ (3f -1)\mu_{\rm Co} + (2-3f)\mu_{\rm M}\\
& & \pm \sqrt{ 8 \mu_{\rm Co}\mu_{\rm M} + ( (3f-1)\mu_{\rm Co} + (2-3f)\mu_{\rm M})^2} ], \nonumber \label{MuCompositeBruggeman}
\end{eqnarray}
where $f$ is the particle volume fraction, $\mu_{\rm Co}$ is the permeability of the cobalt particles, and $\mu_{\rm M}$ is the permeability of the insulating material.

We now show that the formula for the resonance frequency Eq. (\ref{rezz}) remains valid also for composites, no matter what the mixing rule is, under the condition that the material used for the matrix is magnetically inert. This condition is certainly satisfied for our samples. To prove this, let us consider an arbitrary function of the two components $\mu_{M}$ and $\mu_{\rm Co}$,
$\mu_{r} = \mu_{r}[\mu_{M}, \mu_{\rm Co}]$. Then
\begin{equation}
\frac{\partial \mu_{r}}{\partial \omega} = \frac{\partial \mu_{r}}{\partial \mu_{\rm Co}}
\frac{\partial \mu_{\rm Co}}{\partial \omega},
\end{equation}
where we have used the assumption above about the matrix material, namely that its spectrum is flat,
$\partial \mu_{\rm M}/\partial \omega = 0$. Thus, the zeroes of $\partial \mu_{r}/\partial \omega$ will coincide with the zeroes of $\partial \mu_{\rm Co}/\partial \omega$, and  Eq. (\ref{rezz}) can also be used for the composite. This is intuitively very clear: since the Co is the only material in the  composite  that has some magnetic properties (the ferromagnetic resonance in this case), one expects that these properties and only these will be responsible for any structure in the spectra of the composite as well.

\subsection{SQUID magnetometry}
\begin{figure}
\includegraphics[scale=0.85]{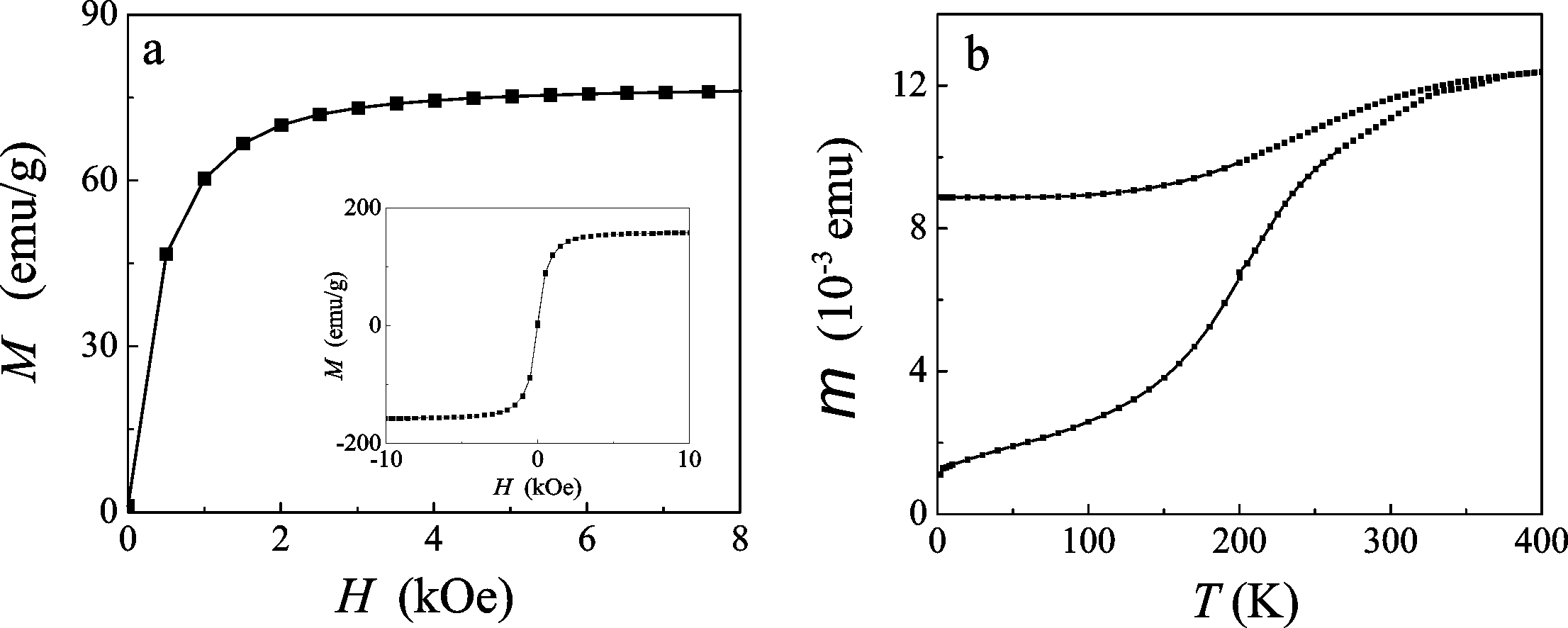}
\caption{The magnetic response of $\epsilon$-cobalt nanoparticle composite measured by a SQUID magnetometer. a) Magnetization $M$ versus applied magnetic field showing a saturation magnetization of about 77 emu/g. The inset picture is the result of another section from the same sample. b) The ZFC-FC curve of the same sample ($m$ denotes the magnetic moment) showing the blocking temperature above room temperature.} \label{Figure2}
\end{figure}

 The effective anisotropy field $K$ and the saturation magnetization $M_s$ can be determined from magnetic measurements done with a superconducting quantum interference device (SQUID) Quantum Design MPMS XL7 magnetometer. These measurements provide as well
 an estimate for the effective anisotropy field $H_{\rm A}$, for which we use the standard result of the Stoner-Wohlfart model
$H_{\rm A} = 2K/\mu_{0}M_s$ (see e.g. Ref. [\onlinecite{Timonen2010}] for a simple derivation). The measurements were performed on small sections from the actual sample, see Fig. \ref{Figure2}a). Each section has a volume of approximately 1 mm$^3$. For convenience, we will express the saturation magnetization in the standard units of magnetization per unit mass (emu/g), which is obtained from the measurements of the weights of all the samples and their volume fractions. When used in the theoretical model, the  magnetization is converted into units of magnetic moment per volume (A/m) by using the density of cobalt, $\rho_{\rm Co}=8.6$~g/cm$^3$.

To determine the effective anisotropy constant $K$, we measure the ferromagnetic-superparamagnetic blocking temperature $T_B$ from the zero-field cooled (ZFC) and field-cooled (FC) magnetization curve. Fig. \ref{Figure2}b shows a typical ZFC/FC curve of a $\epsilon$-cobalt nanoparticle composite. The measurement was realized by firstly cooling the small piece of sample to 2 K without an external magnetic field for a ZFC measurement. At the end, the sample was freezed with no net magnetization, due to the random magnetization at room temperature. Next, a small field (100 Oe) was applied, and the magnetization ($M$) of the sample was measured at different temperature from 2 to 400 K. As the temperature increases, more particles go from the ferromagnetic (blocked) to the paramagnetic phase, and align with the applied field. The magnetization reached a maximum when the blocking temperature was reached.

\subsection{Microwave measurements}

Standard experiments on FMR such as those analyzed by Kittel were done by monitoring the on-resonance response of a microwave cavity while sweeping the external magnetizing field. In contrast, our setup is broadband, allowing to monitor the response at all frequencies. The complex magnetic permeability of the cobalt samples were measured over a frequency range between 1 and 12 GHz by a transmission and reflection method. Prior to the measurements, a sample was inserted inside the coaxial line (7-mm precision coaxial air line) which was placed between the poles of an electromagnet (the axis of the line being perpendicular to the magnetic field). The line was connected to the ports of a vector network analyzer (VNA) by Anritsu 34ASF50-2 female adapters. The transmission/reflection signals were measured and used as the input parameters of the reference-plane invariant algorithm to obtain the complex permittivity and permeability \cite{Chalapat2009}. Similar results are obtained using the short-cut method \cite{Vepsalainen2013}.

\section{Results and Discussion}
\label{externalfield}

\begin{figure}
\includegraphics[scale=0.85]{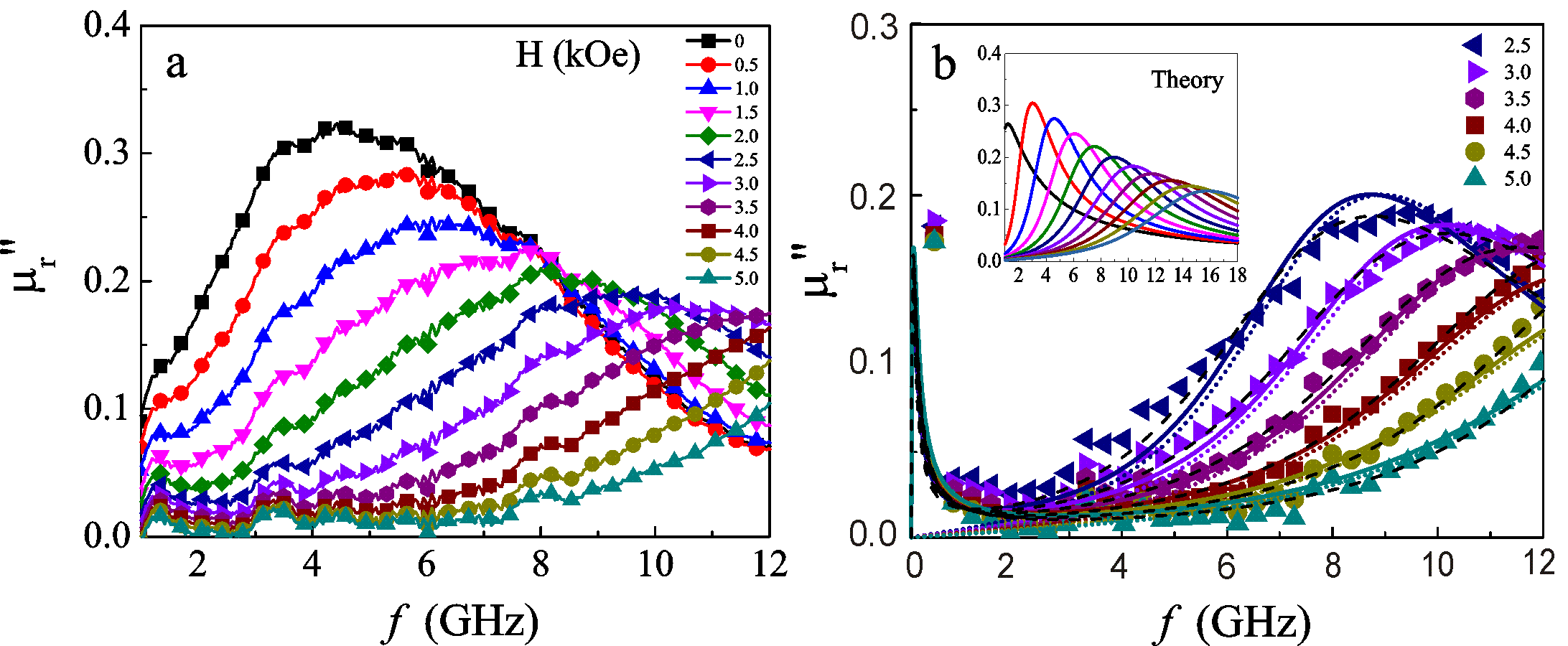}
\caption{a) Imaginary magnetic permeability of a composite of polystyrene (PS) and $\epsilon$-cobalt particles, at a volume fraction of 10\%. b) Theoretical predictions: LLG-Kittel (dotted lines) and Debye/LLG-Kittel (solid and dashed lines), shown in comparison with the experimental data (discrete symbols) at high magnetic fields ($\geq 2.5$ kOe). The average particle diameter was $d = 16$ nm. Inset: theoretical prediction of LLG-Kittel for a wider range of field values.}\label{Figure3}
\end{figure}

\paragraph{Measurements in externally applied magnetic fields.}
We start by presenting the effects of an external magnetizing field on the FMR spectra of composites made with spherical $\epsilon$-cobalt nanoparticles synthesized by the hot-injection method \cite{Puntes2001}. We call SET-1 this first set of samples (and similar notations will be used for the rest of the sample sets, see Table \ref{Table}). We measured the relative complex magnetic permeability ($\mu_r = \mu_r'-i\mu_r''$) of these samples over a wide range of frequencies for various non-zero external static fields $H$. The material exhibits a broad resonance peak around 4 GHz in a zero external field, see Fig. \ref{Figure3}a. The application of an external magnetic field causes the resonance to shift towards higher frequencies, accompanied by a reduction of the magnetic loss peak. Interestingly, in the small field regime (below 1.5 kOe), the magnetic absorption remains almost independent of the field at frequencies above 8 GHz.

Fig. \ref{Figure3}b presents a comparison between the theoretical models and experimental results. The theoretical spectra ($\mu_r''$) were calculated by combining the phenomenological models: LLG-Kittel \cite{LandauLifshitz,Kittel1948} and Debye/LLG-Kittel \cite{Fannin1997} with the Bruggeman effective
medium model \cite{Bruggeman1935}. The simulation was done with a particle volume ratio of 0.1, and an average particle diameter $d$ of 16 nm. The magnetizing field
corresponds to the experimentally-measured magnetic field at the sample (using a calibrated gaussmeter)
and ranges from 2.5 to 5.0 kOe. The dotted lines show the prediction of the LLG-Kittel model,
Eq. (\ref{linear}), Eq. (\ref{LLGImMuParticle}) and Eq. (\ref{MuCompositeBruggeman}),
with parameters $M_s = 64$ emu/g, $T_B = 400$ K, and $\alpha = 0.37$.
The solid and dashed lines show the prediction of the Debye/LLG-Kittel model,
Eq. (\ref{linear}), Eq. (\ref{LLGImMuParticle}), Eq. (\ref{DebyeMu}) and Eq. (\ref{MuCompositeBruggeman}).
The Debye model (parallel susceptibility) was calculated by setting $\alpha_{\parallel} = 5\times 10^{-4}$ and $\chi_\parallel(0) = 10^4$. The solid lines present the theoretical prediction with $M_s = 74$ emu/g, and $\alpha = 0.37$. An even better fitting with the data can be obtained (dashed lines) by allowing for a small magnetic-field dependence of the damping $\alpha$: 0.44 ($H = 2.5$ kOe), 0.40, 0.36, 0.32, 0.28 and 0.24 ($H = 5$ kOe) with $M_s = 84$ emu/g.

The analysis presented in Fig. \ref{Figure3}b suggests that both LLG-Kittel and Debye/LLG-Kittel models can approximately predict the FMR spectra in the regime where the magnetizing field is large. The LLG-Kittel (dotted lines) predicts that the imaginary part of magnetic permeability $\mu''$ should go to zero at very low frequencies. In the experiment however we see that the decrease in the permeability $\mu''$ does not continue indefinitely at low frequencies, but instead starts to increase again below 2 GHz. This large absorption at the low-frequency range is governed by non-resonant relaxation processes, which can be estimated by the Debye/LLG-Kittel model (solid lines).

\begin{figure}
\includegraphics[scale=0.4]{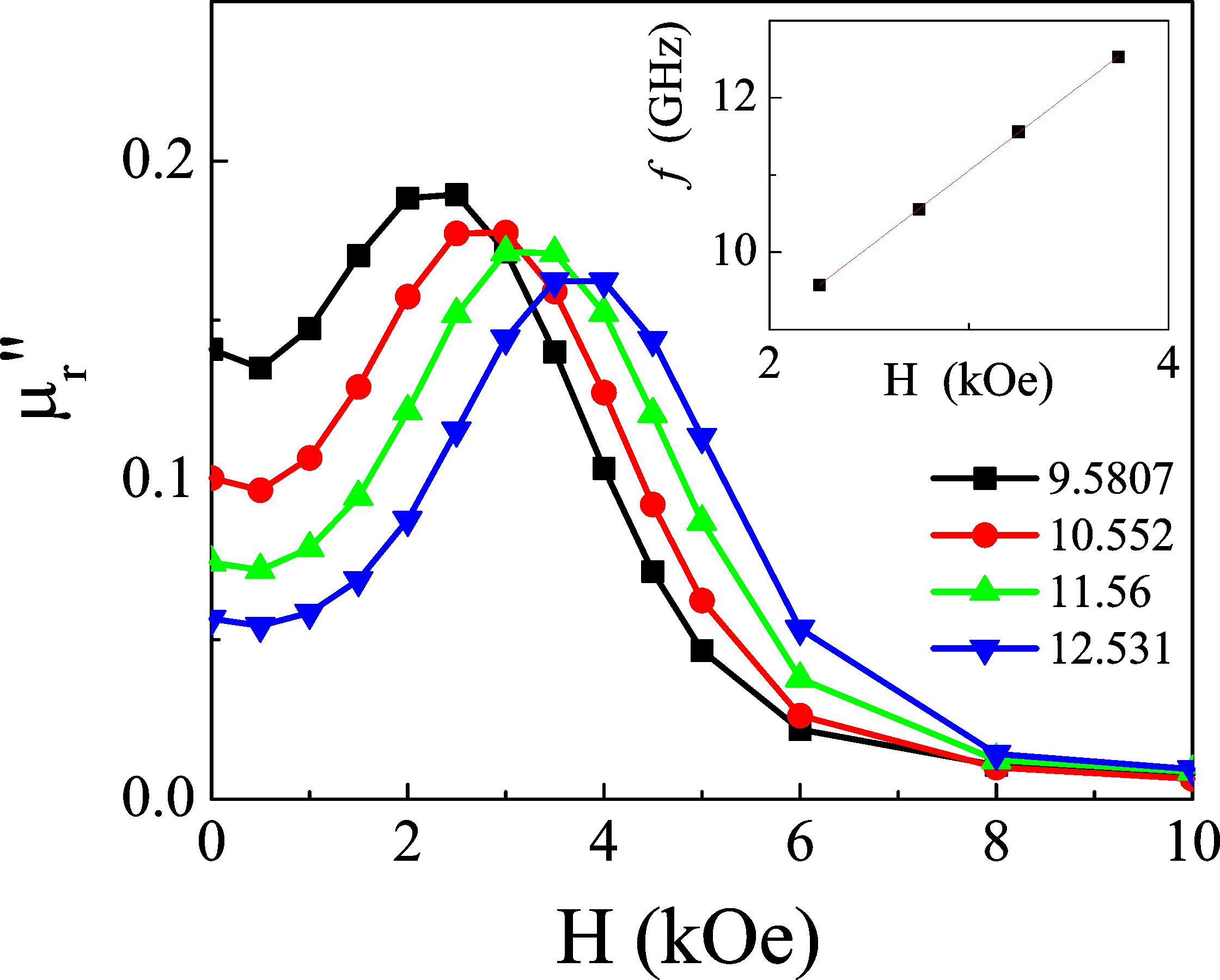}
\caption{The imaginary part of magnetic permeability as a function of the applied magnetic field $H$. Each color represents the peak at one microwave frequency in the external field domain. The measurement shows that the resonance shifts towards a higher magnetic field when the microwave frequency is increased. The inset picture shows the linear dependence between the resonance frequency $f_r$ of the composite and the external magnetic field $H$. \label{Figure6}}
\end{figure}

Next, we present the permeability data in the magnetizing-field domain. Four sets of data that show nearly-Lorenzian curves are plotted in Fig. \ref{Figure6} to show the FMR peaks. The absorption spectra shown in Fig. \ref{Figure6} provide a direct proof of the linear relation between the resonant microwave frequency and the magnetizing field (inset graph of Fig. \ref{Figure6}).
According to Eq. (\ref{rezz})
the slope of the $f_r-H$ predicted by the LLG-Kittel model is $\gamma/2\pi\sqrt{(1+\alpha^2)} = 2.62$ GHz/kOe for $\alpha = 0.37$ and $\gamma=\gamma_{e}$. From the data, we can estimate an average slope of $2.4$ GHz/kOe, remarkably close to the ideal value.


\begin{figure}
\includegraphics[scale=1.4]{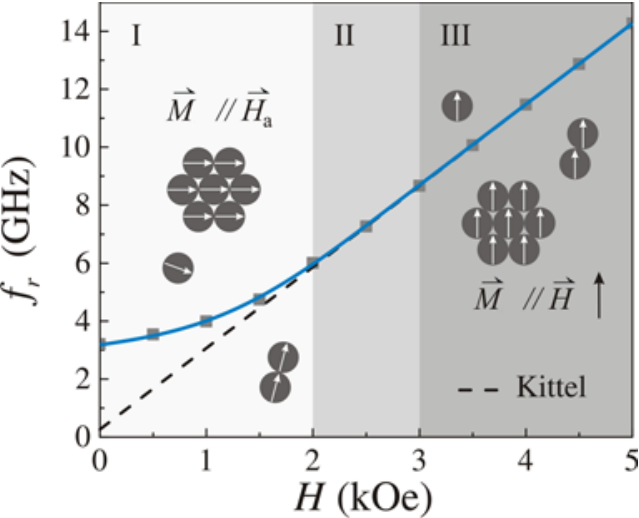}
\caption{Diagram showing the relation between the resonance frequencies ($\omega_0$) and the applied magnetizing field ($H$). The resonant frequencies are obtained for each applied external field from the fitting with the LLG/Kittel-Bruggeman equations, and plotted as square markers.
In the low field regime (I), the magnetization of the particles is random and aligns along local anisotropy fields (denoted by $H_a$). When chains or clusters are present in the composite, the local fields cause an increase of the FMR resonance frequency (continuous line). In the intermediate regime (II) and in especially in the high field regime (III), the magnetization of the particles is directed along the external magnetizing field, and Kittel's FMR theory predicts correctly the resonance absorption spectra (dotted line).}\label{Figure4}
\end{figure}

At low magnetic fields, discrepancies with respect to the LLG-Kittel theory start to appear.
Fig. \ref{Figure4} includes the low-field dependence of the ferromagnetic resonance, showing that the linear relation Eq. (\ref{rezz}) valid at high fields in region (III) does not hold anymore in region (I). Extrapolating the linear dependence $\omega_{r} = \gamma (H + \tilde{H})$ of region (III) to low field values
and estimating $\tilde{H}\approx H_{\rm A}$
still yields a good estimate for the effective magnetic anisotropy constant $K$ (energy per unit volume) of the order of $10^4$ J/m$^3$, with $M_s=84$ emu/g
as used in the Debye/LLG-Kittel model (dashed lines) in Fig. \ref{Figure3}b. However, in the low field (I) regime, the resonance occurs at a slightly higher frequency.
To understand the origin of this frequency shift, we imaged the $\epsilon$-cobalt nanoparticles with a transmission electron microscope (TEM), and found that they tend to agglomerate in clusters, see Fig. \ref{Figure5}. When the particles aggregate, the distance between them might be small enough to create a sizable magnetic interaction. In each cluster,
the dipole interactions between particles will
generate additional local fields $H_{a}$, which are not accounted for in the noninteracting model, see {\it e.g.} Eq. (\ref{linear}). These interaction effects become dominant in the regime of small magnetic fields (I).
\begin{figure}
\includegraphics[scale=1.5]{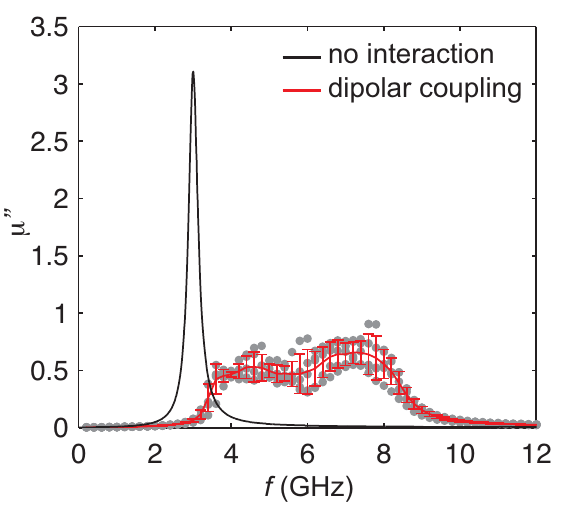}
\caption{Simulated imaginary part of the permeability for two monodisperse $\epsilon$-Co nanoparticles (10 nm) in random 3D aggregates with surface-to-surface separation of magnetic cores of 5 nm and damping $\alpha =0.05$.
\label{Figure_dipolar}}
\end{figure}
This magnetostatic dipolar interaction between particles gives an additional component to the total anisotropy, now proportional to $M_s/d^3$ where $d$ is the interparticle spacing. The effect of this interaction can be obtained by solving numerically the LLG equation using a micromagnetic simulation package \cite{Fischbacher2007,Timonen2012} for two particles. When the dipolar coupling is not enabled the free particle resonance is obtained. When dipolar coupling is turned on, the resonance shows significant broadening and shifting toward higher frequencies due to the formation of collective resonance modes (see Fig. \ref{Figure_dipolar}). Although this is a simplified model with only two particles, the result supports the experimental observation that the resonance  frequency is shifted to higher values (see Fig. \ref{Figure4}). This is also in agreement with the result of Ref. [\onlinecite{Buznikov}]. The appearance of collective behavior due to dipolar magnetic percolation and the formation of correlated agglomerates has been also observed  experimentally by magnetic force microscopy in two-dimensional Co layers \cite{PuntesAlivisatos2004}.

\begin{figure}
\includegraphics[scale=0.27]{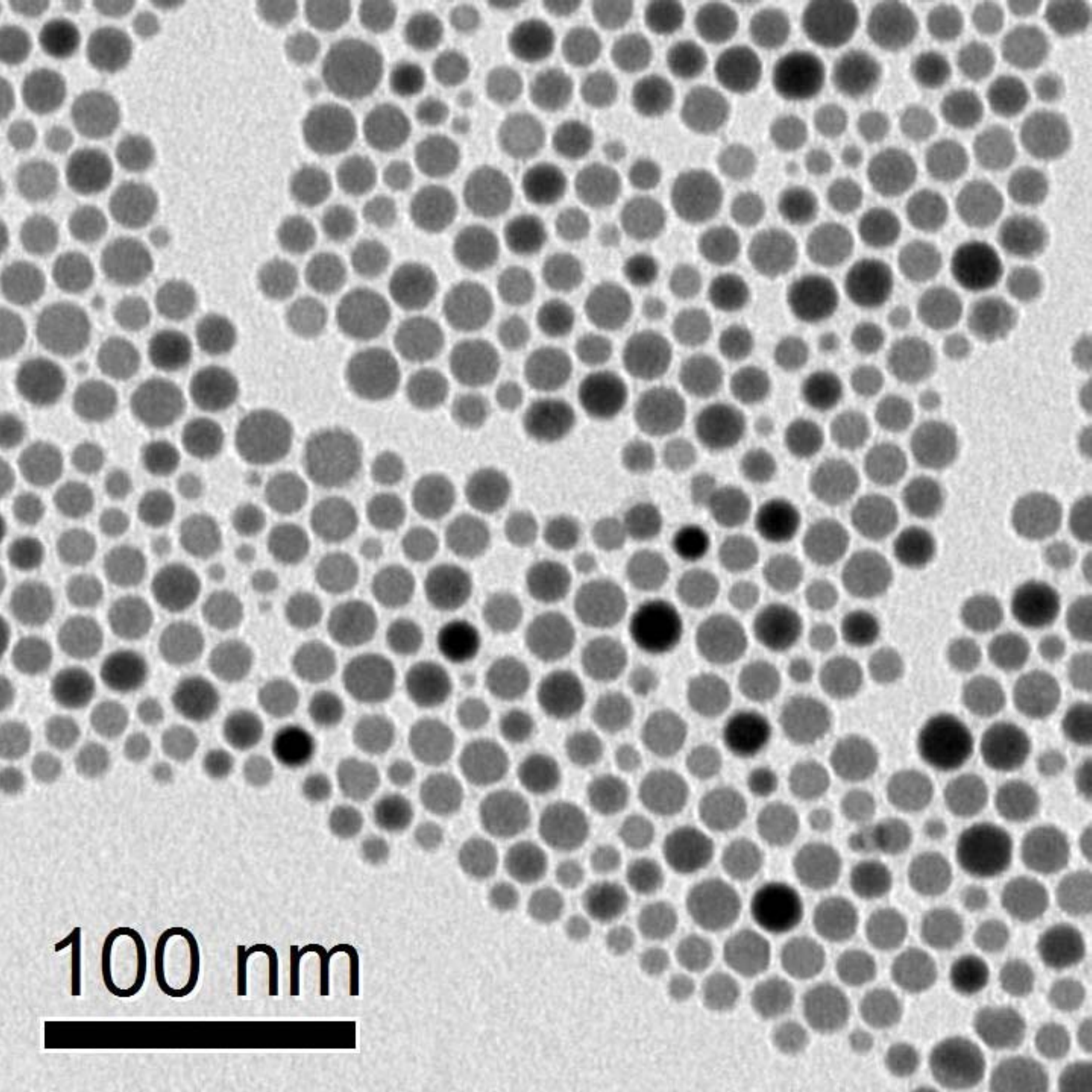}
\caption{TEM  image of Co nanoparticles within SET-1 before mixing with polystyrene. The average diameter is in the order of 16 nm (with most of the particles between 10 and 20 nm). \label{Figure5}}
\end{figure}


\paragraph{Measurements at zero magnetic field.}

For technological application it is not always possible to apply large magnetic fields, thus one should provide a systematic classification of the composites. Since we cannot control the aggregation patterns of the particles, we have chosen instead to modify the structure, distribution, and morphology of  the constituent nanoparticles. The measurements show that this does not produce qualitative changes in the spectra, and the results are compared with the predictions of the LLG-Kittel complemented with the Bruggeman and Maxwell-Garnett mixing rules. These measurements are also very useful for technological applications, where precise knowledge of the electromagnetic parameters of these composite materials is needed. The particle-level properties of these composites are summarized in Table \ref{Table}.

\begin{table*}[ht]
\centering
\begin{tabular}{|c|c|c|c|c|}
\hline\hline
sample set & volume fraction,  f [\%] & diameter, $w \pm \sigma$ [nm] & dispersion, $100 \cdot\sigma/w$ [\%]& synthesis method\\ \hline
\hline\hline
SET-1 & 10 \%              & 16 $\pm$ 5 nm & 31\% (polydisperse) & hot injection \\
\hline
SET-2 & 4\% and 11 \% & 13.9 $\pm$ 0.6 nm & 4\% (monodisperse) & heating-up \\
\hline
SET-3A & 4.3\% & 8.6   $\pm$ 1.4 nm & 16\% (polydisperse) & heating-up \\
SET-3B & 1.9\% & 27.7  $\pm$ 2.7 nm & 10\% (intermediate) & heating-up \\
SET-3C & 1.6\% & 32.1  $\pm$ 7.9 nm & 25\% (polydisperse) & heating-up \\
\hline
SET-4A & N/A & 5 $\pm$ 0.6 nm & 12\% (intermediate)  & hot injection \\
SET-4B & 3.3\%, 6.5\%, and 6.9\% & 8 $\pm$ 0.8 nm & 10\% (intermediate) & hot injection \\
SET-4C & 2.9\% and 6.7\% & 11 $\pm$ 0.9 nm & 8\% (intermediate) & hot injection \\
SET-4D & 7.7\% and 8.5\% & 27 $\pm$ 3.5 nm & 13\% (intermediate) & hot injection \\
\hline
\end{tabular}
\caption{Systematic of the nanoparticle properties of various samples fabricated and measured. Note: very small particles were observed to be present in all solutions; they were not included in the calculation of the average diameter, and they are therefore not subsequently included in the estimations of the effective anisotropy, as these particles are magnetically inert.
}
\label{Table}
\end{table*}

\begin{figure}[ht]
\includegraphics[scale=0.27]{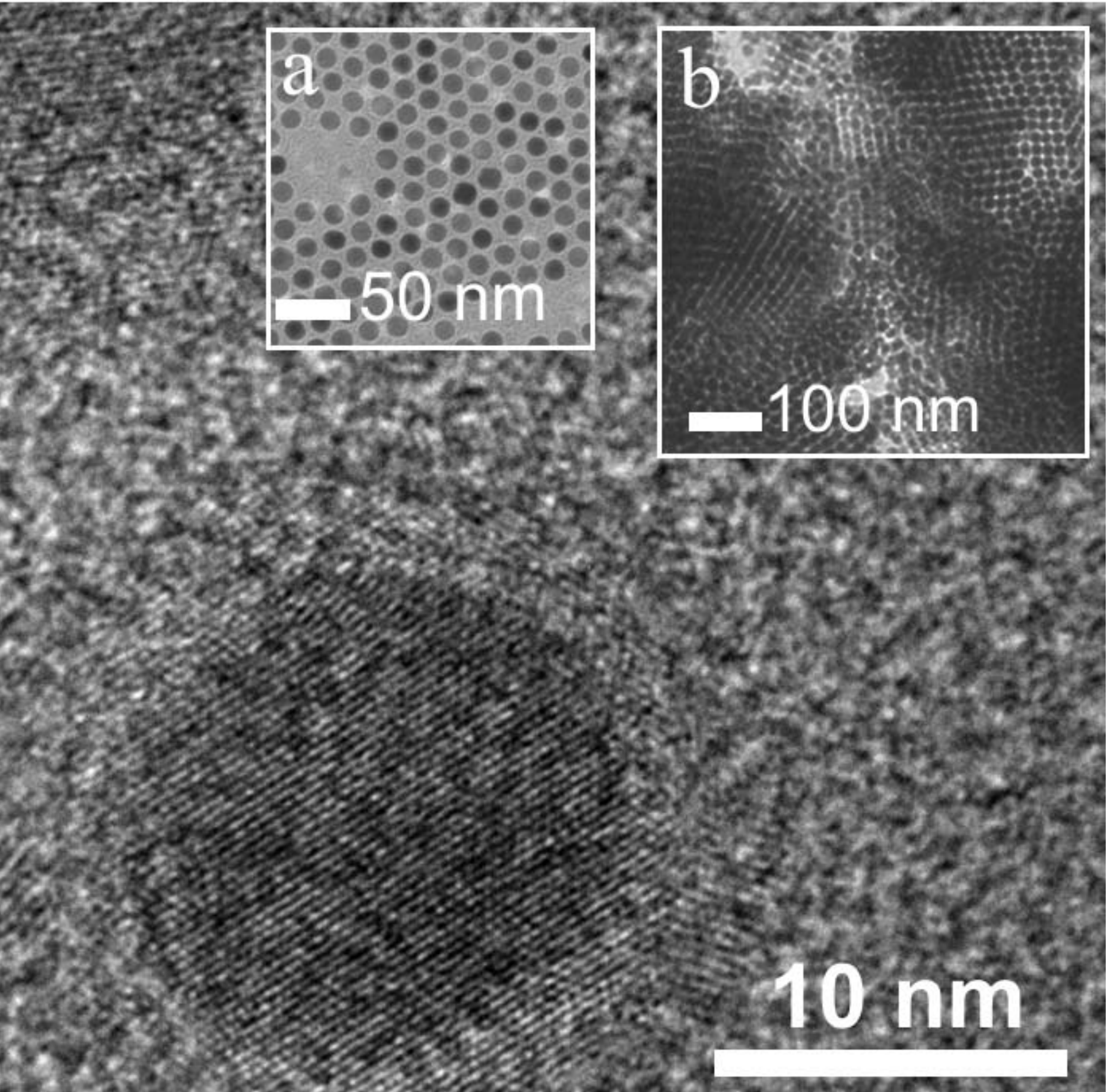}
\caption{TEM images of cobalt nanoparticles in SET-2. The main figure shows a single crystalline core inside a particle with an average diameter of about 14 nm. Note the high quality of the particle: a single crystal, with no grain boundaries, no dislocations or other defects.
a) The first inset shows the particles after being left dry on a TEM grid. The size distribution of the particles is very narrow (almost equal size). b) The second inset shows the arrangement of particles inside a cluster in the composite; superlattices (closed-pack array of particles) are seen in many parts of the cluster. \label{Figure7}}
\end{figure}

\begin{figure}[ht]
\includegraphics[scale=0.9]{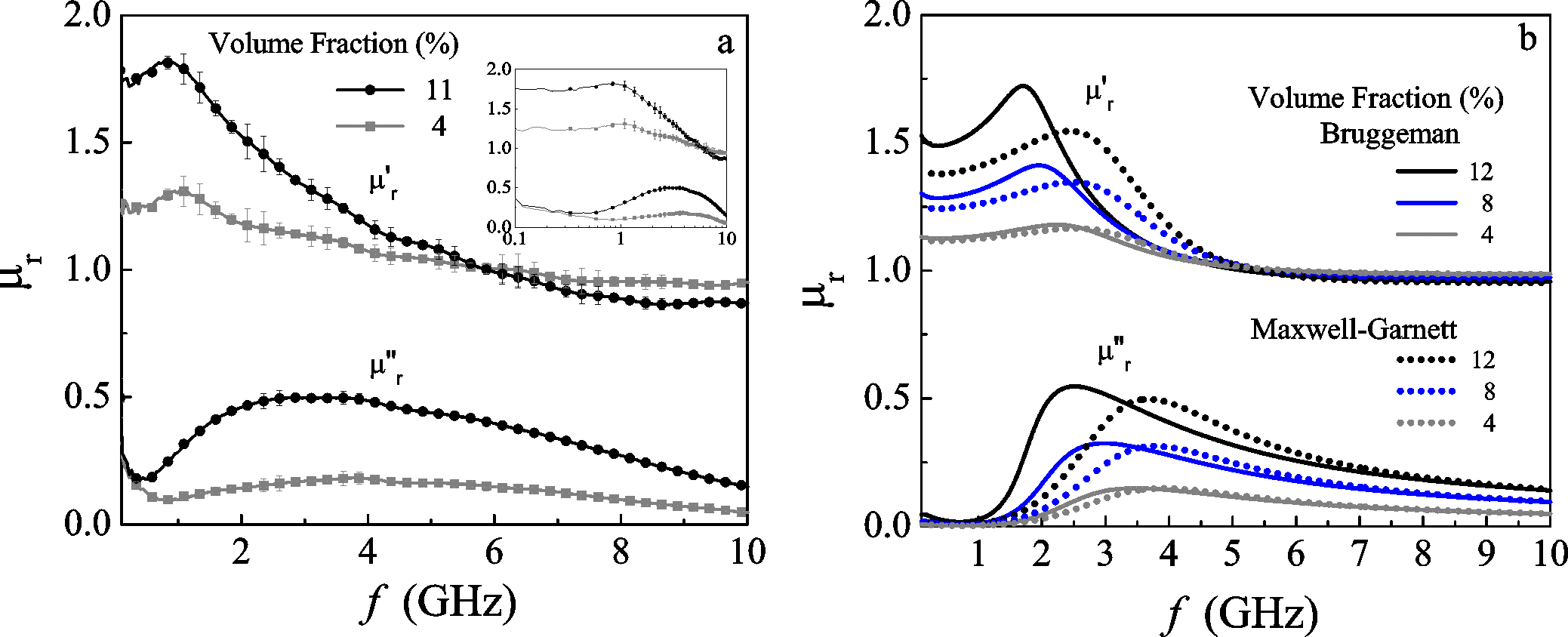}
\caption{a) Complex magnetic permeability for SET-2. b) Theoretical spectra calculated using Maxwell-Garnett and Bruggeman effective medium models with the volume fraction of 0.11. The LLG/Kittel modelling was done with $d$ = 14 nm, $M_s$ = 150 emu/g, $T_B$ = 400 K, $\alpha$ = 0.12, and $\tilde{H} = H_{\rm A} + 0.3~{\rm kOe}$ (or $\mu_{0}\tilde{H} = \mu_{0}H_{\rm A} + 30~{\rm mT}$), with
$H_{\rm A}$ estimated from $H_{\rm A} = 2K/\mu_{0}M_s$ and
$K \approx k_{B}T_{B}/V$, where $V$ is the average particle volume. The Debye part was calculated using $\alpha_\parallel = 2\times 10^{-3}$ and $\chi_\parallel(0) = 10^{4}$. \label{Figure8}}
\end{figure}

The monodispersed $\epsilon$-cobalt particles of SET-2 are fabricated by the heating-up chemical synthesis method \cite{Timonen2011}. The TEM images of these particles show that the particles have an average diameter of 14 nm, see Fig. \ref{Figure7}.
We measured the permeability spectra of two samples with different volume fractions. The results are shown in Fig. \ref{Figure8}a. We notice that increasing the volume fraction not only increases the magnitude of $\mu''_r$, but also shifts the resonant peak towards lower frequencies. The calculations based on effective medium models show that the Bruggeman equation also predicts that the resonance absorption ($\mu''$ peak) shifts towards lower frequencies when the volume fraction is increased (see Fig. \ref{Figure8}b). In Fig. \ref{Figure8}, the theoretical graph was determined with an effective blocking temperature of 400 K, reproducing correctly the resonant peak around 3 GHz.
 An estimation of the effective magnetic anisotropy $K$ from the exponent of the N\'eel-Arrhenius law $T_B \approx KV/k_B =$ 400 K yields $K \approx$ 3.84$\times 10^3$ J/m$^3$ = 3.84$\times 10^4$ erg/cm$^3$. This effective magnetic anisotropy is lower than the anisotropy constant of $\epsilon$-cobalt extracted from magnetic measurements \cite{PuntesKrishnan2001}. This is an expected limitation of our simplified model, in which a single effective superparamagnetic blocking temperature is used as a parameter to fit the permeability spectra. For our samples, different parts of the composite may have different blocking temperature, depending on how clusters are formed. For example, it is known that
in ferromagnetic films the blocking temperature rises when the grains coalescence, and also when the magnetic domains grow \cite{Frydman1999}.


\paragraph{Particle-size and morphology effects.}

\begin{figure}
\includegraphics[scale=0.30]{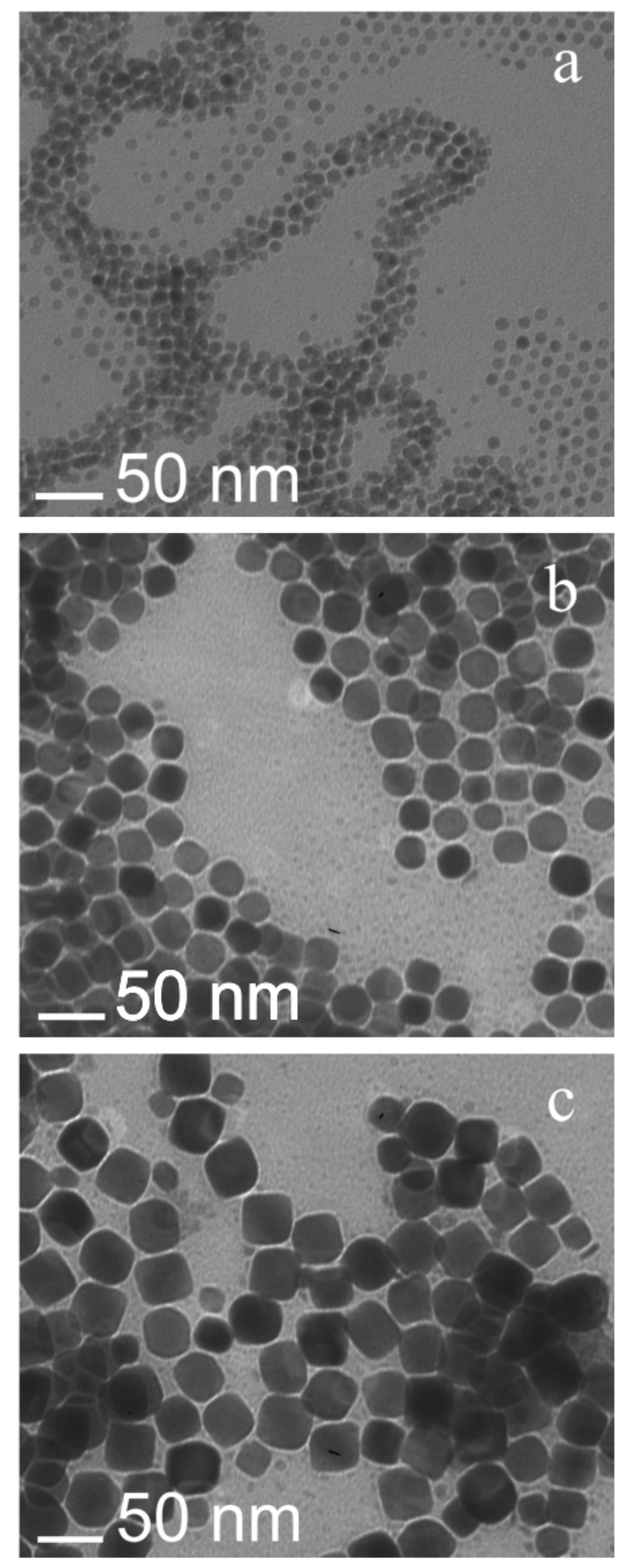}
\caption{TEM images of $\epsilon$-cobalt nanoparticles in SET-3. a) SET-3A contains particles with the smallest average diameter (8.6$\pm$1.4 nm) showing some aggregation in the form of chains b) SET-3B contains particles with the average diameter of 27.7$\pm$2.7 nm, and c) SET-3C contains particles of the average diameter of 32.1$\pm$7.9 nm.
\label{Figure10}}
\end{figure}

Next, we investigate how the particle size and particle morphology affects the microwave absorption properties of a composite made from $\epsilon$-cobalt particles. First, we study three new composites (SET-3A, SET-3B, SET-3C), from a third set of samples (SET-3). The TEM images of the particles within these composites are shown in Fig. \ref{Figure10}. We see that sample-A contains spherical nanoparticles with the average diameter of $8.6\pm 1.4$ nm, while sample-B and sample-C represent the composites made from bigger particles: $27.7\pm 2.7$ nm and $32.1\pm 7.9$ nm, respectively. The images also suggest that the heating-up synthesis method \cite{Timonen2011} can produce $\epsilon$-cobalt nanoparticles exhibiting facet-structures if the particle sizes are bigger than 25 nm.

\begin{figure}
\includegraphics[scale=0.45]{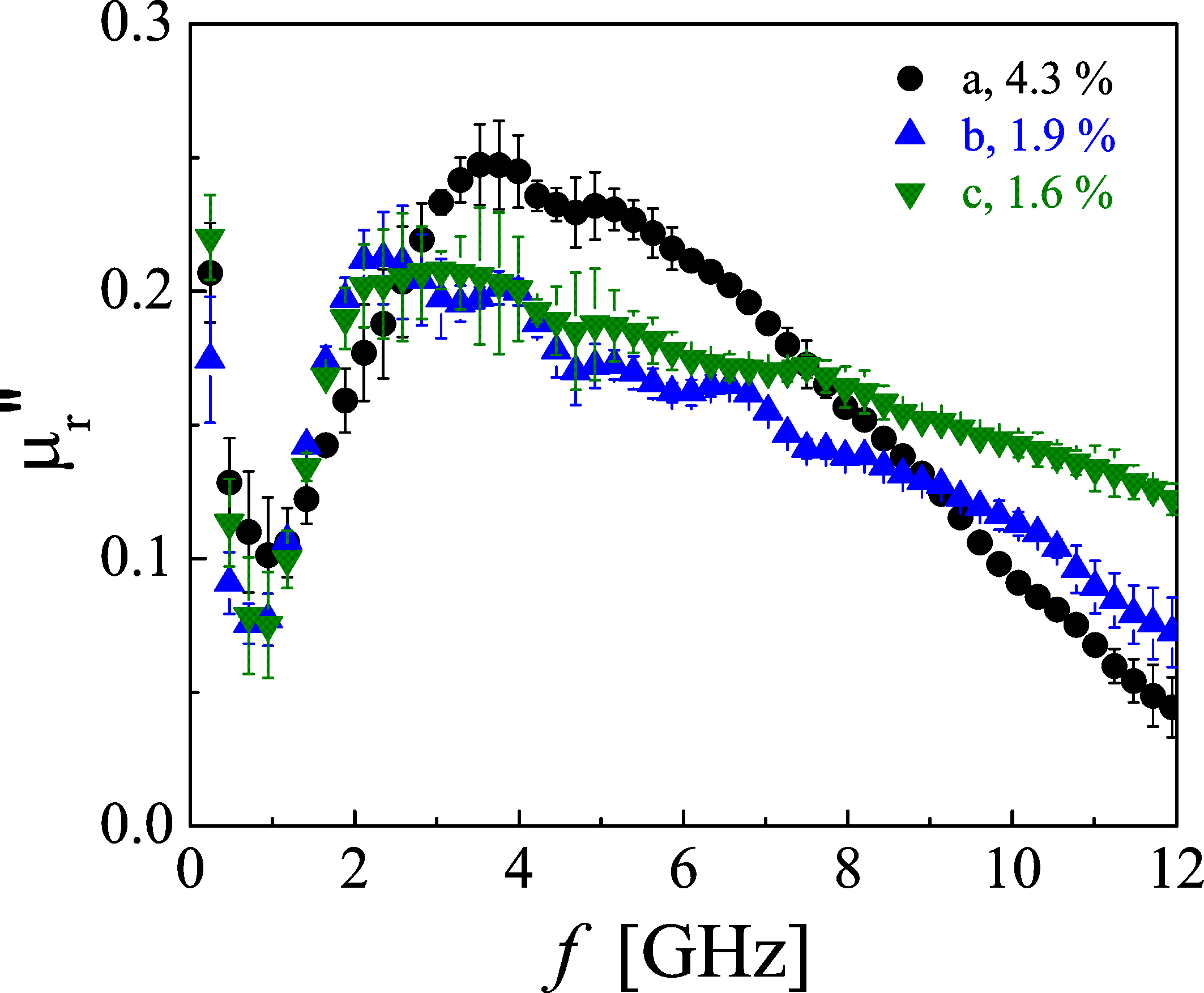}
\caption{The imaginary part of the magnetic permeability of (a) SET-3A, (b) SET-3B, and (c) SET-3C. \label{Figure11}}
\end{figure}

\begin{figure}
\includegraphics[scale=1.2]{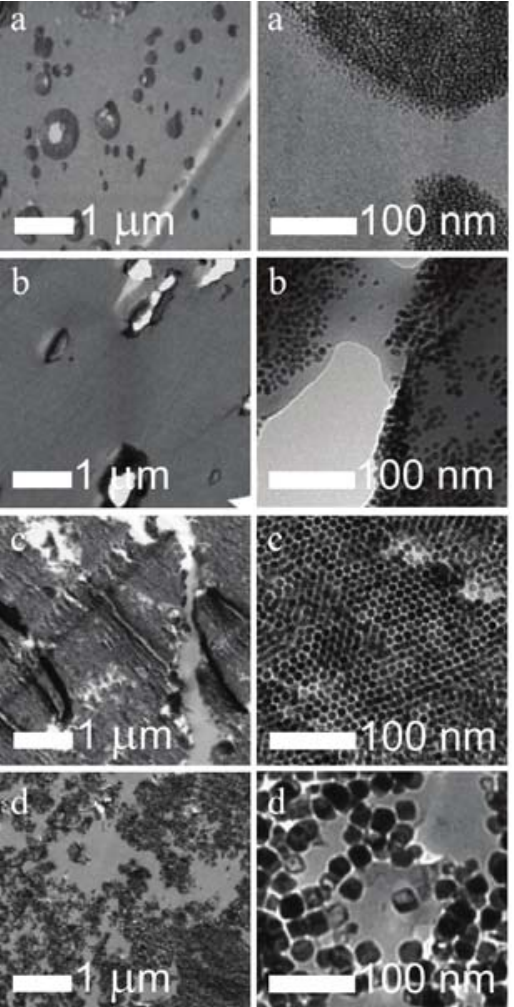}
\caption{TEM images of $\epsilon$-cobalt nanoparticles in SET-4. a) SET-4A contains nanoparticles with average diameter of 5 nm. b) SET-4B contains nanoparticles with average diameter of 8 nm, c) SET-4C contains nanoparticles with average diameter of 11 nm, and d) SET-4D contains bigger facet-particles with average size of 27 nm.\label{Figure12}}
\end{figure}

The magnetic permeability of these samples (Fig. \ref{Figure11}) demonstrate that faceted particles (non-spherical) exhibit a similar FMR absorption peak as the smaller spherical particles. The absorption is highest between 2 and 5 GHz. Note that the samples that are made from bigger (25-40 nm) $\epsilon$-cobalt particles tend to have a wider absorption bandwidth. TEM imaging confirms that nanoparticles larger than 25 nm form facet structures, see Fig. \ref{Figure12}d. These non-spherical particles (Fig. \ref{Figure10}b-c and Fig. \ref{Figure12}d) cause their composites to exhibit a small absorption at 1 GHz ($\mu''_r$ is less than 0.1) and a large absorption (resonance) over a wide frequency range above 1.5 GHz.


\begin{figure*}
\includegraphics[scale=0.85]{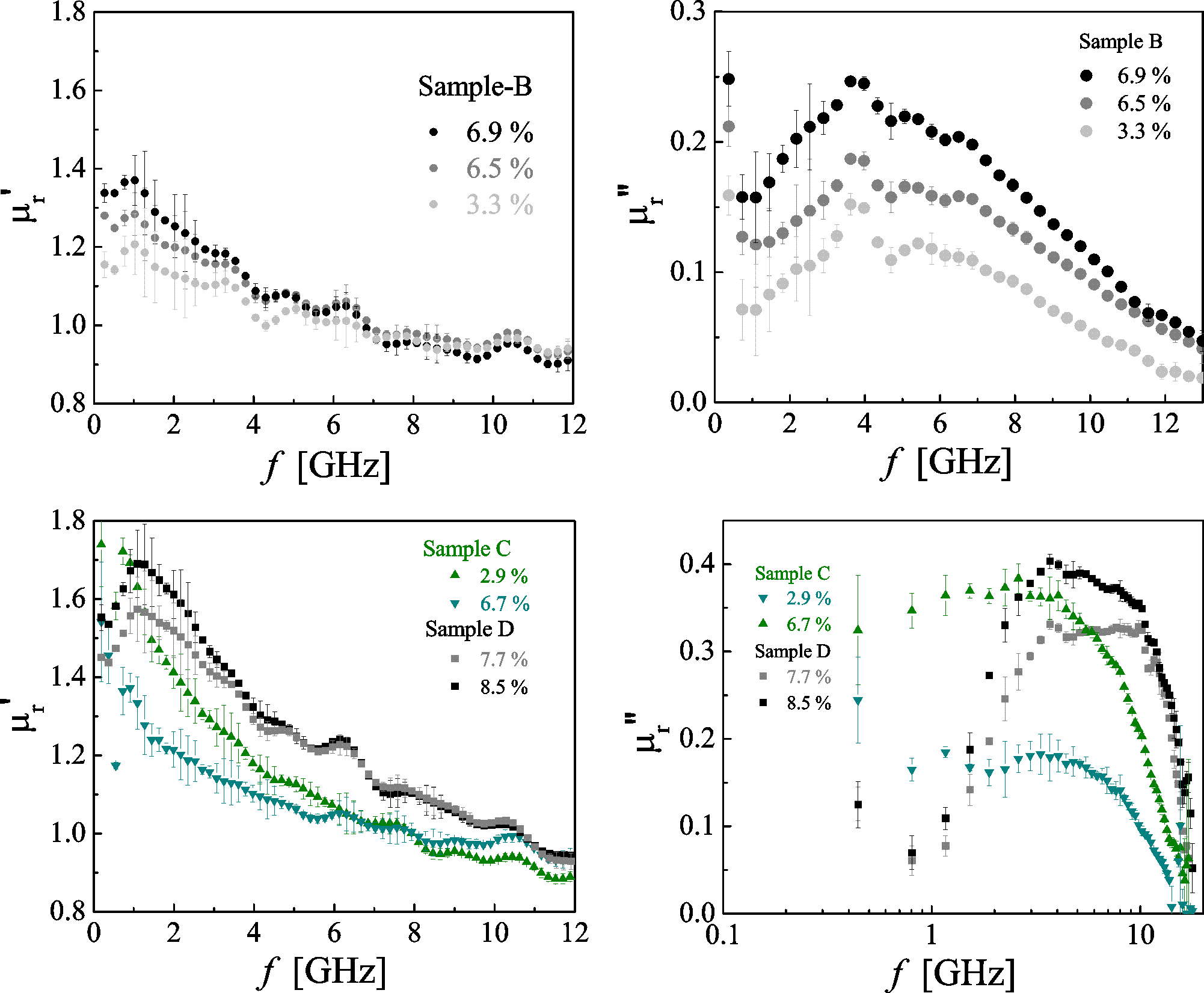}
\caption{The real and the imaginary part of the magnetic permeability for SET-4B, SET-4C, and SET-4D. \label{Figure15}}
\end{figure*}



From the measurements presented in Fig. \ref{Figure15}, we notice that composites made of low-dispersion 11-nm particles (SET-4C), see Fig. \ref{Figure12}c, have relatively large absorption at frequencies around 1 GHz, compared to the absorption given by larger-size particles (27 nm, SET-4D) see Fig. \ref{Figure12}d. This may be due also to the large fraction of closed-pack structures and aggregates. We also found that smaller particles, with an average diameter of 5 nm (SET-4A in Fig. \ref{Figure12}a), do not exhibit significant absorption in this range.

The magnetic permeability of SET-4 shows that a composite made from particles with diameter smaller than 8 nm is a weak absorber compared to composites made from bigger (11 to 35 nm) particles (Fig. \ref{Figure15}). This observation suggests that the microwave absorption of SET-3A (Fig. \ref{Figure10}a) is associated with the FMR of minority particles that have bigger sizes (those which form chain structures). In principle, chains can form during the wet-chemical synthesis because of the presence of magnetic dipole interaction. Aggregation can change the anisotropic energy and so cause the unexpected (small) change of microwave absorption spectra. From Fig. \ref{Figure15}, we find that the composites with the average particle diameter of 27 nm exhibit high permeability ($\mu'_r \approx 1.7$) at 1 GHz, with the magnetic loss $\mu''_r$ of less than 0.1.

In consequence, for the applications in the area of low-loss devices/materials, the experiments show that it is sufficient to use $\epsilon$-cobalt facet-nanoparticles with sizes of about 20 nm. For technological applications in the area of microwave absorbers we find that the absorption spectra are also sensitive to the way particles arrange themselves inside the composite materials. The formation of superlattices in monodispersed particles could result in large absorption over a broader bandwidth. Ideally, one should also aim at controlling the particle-particle interaction (distance) as well as at creating regular (periodic) structures, which can be in principle done by novel self-assembly and polymerization techniques. One possibility would be to use in the synthesis a surfactant with an acid group at one end and a double bond at the other end (for example docos-21-enoic acid), in order to obtain particles with a double bond functionality. After mixing the particles with styrene and polymerization, the interparticle distance will be limited to the same value for all particles.

\begin{figure}[ht]
\includegraphics[scale=0.4]{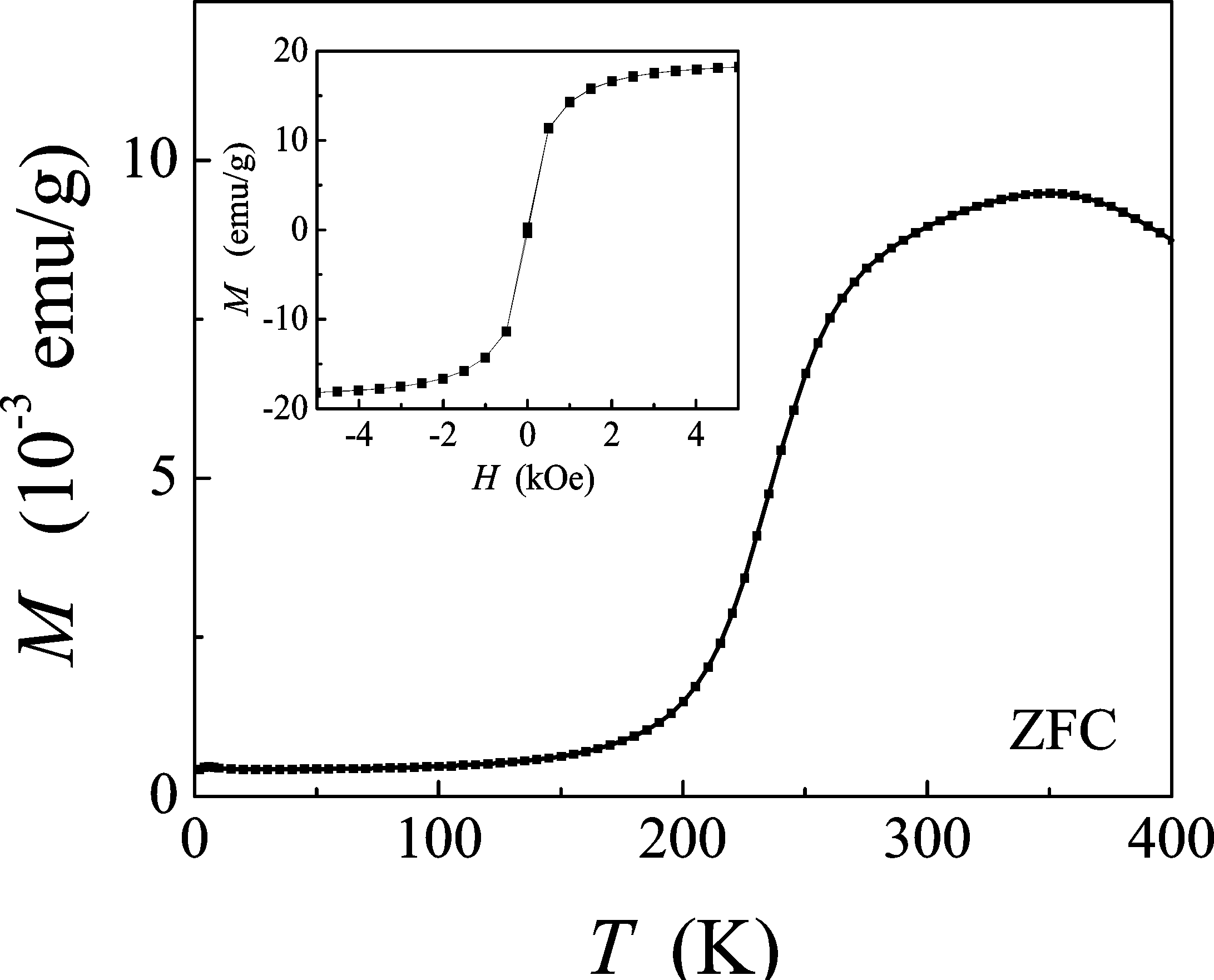}
\caption{The ZFC-FC curve of SET-2, five months after synthesis. The average particle diameter is 14 nm. The inset figure shows that the saturation magnetization is below 20 emu/g. \label{Figure9}}
\end{figure}

\paragraph{Effects of aging.}
Due to the natural oxidation of cobalt particles, the magnetic permeability of a composite sample will decrease in time. This reduction is associated with the reduction of saturation magnetization. Fig. \ref{Figure9} shows the magnetization (zero-field-cooled curve) of SET-2 measured by SQUID magnetometry five months after the synthesis. We observe a change in $T_B$ from 400 K to 350 K. According to the LLG-Kittel theory, this will not change the permeability spectra that much. However, the decrease of saturation magnetization $M_s$ due to oxidation can cause a significant reduction of $\mu'$ values.

\section{Conclusion}
\label{conclusion}

We have experimentally studied the wideband microwave absorption of composite materials made from $\epsilon$-cobalt nanoparticles and polystyrene. The experiments show that the randomly-oriented spherical nanoparticles inside a composite induce ferromagnetic resonant absorption at microwave frequencies (1 to 12 GHz). Composites consisting of $\epsilon$-cobalt nanoparticles within the size range between 8 and 35 nm, either monodispersed or polydispersed, exhibit resonant absorption which peaks in the frequency range between 2 and 6 GHz. Particles of smaller sizes, especially the ones below 5 nm, do not have significant response to the microwave field in this frequency range. The permeability spectra is well described, especially at high-magnetizing fields, by the LLG-Kittel equation and the effective medium model (Bruggeman or Maxwell-Garnett). At zero-magnetizing field, the LLG-Kittel equation and the effective medium model predicts magnetic resonance at lower frequencies compared to the experimental values. We analyze this finding qualitatively by dividing the magnetic response of nanoparticles into three regimes, namely low-field, intermediate-field, and high-field. In case of polydispersed $\epsilon$-cobalt particles with sizes of about 10 nm, the high field regime begins at a magnetizing field of about 3 kOe.


For technological applications, our results demonstrate that the absorption spectra of composites made with either monodispersed and polydispersed particles can be tuned by the application of an external magnetic field. Also, for high-$\mu$ low-loss applications around 1 GHz a good choice are faceted (non-spherical) particles with sizes of about 27 to 35 nm. Medium-size monodispersed spherical particles (10-20 nm) are good as microwave absorbers, particularly if aggregates are formed. Our measurement results show that in this case the absorption capability of the material is increased, thus decreasing the amount of material needed for fabrication of absorbers and filters. This suggests an alternative route to functionalizing these materials through the control of the arrangement of particles, which could be used for engineering tunable microwave circuits and RF components without the need of a magnetic field.


\section*{Acknowledgements}
 We thank M. Sarjala, W. Skowronski, S. van Dijken, A. Savin, and J. Seitsonen for discussions and technical advice. This project was supported by Thailand Commission on Higher Education, the Academy of Finland (projects 118122, 141559 and 135135), and by the Finnish Funding Agency for Technology and Innovation (TEKES) under the NanoRadio project. This work was done under the Center of Excellence "Low Temperature Quantum Phenomena and Devices" (project 250280) of the Academy of Finland, and it used the Aalto University Nanomicroscopy Center (Aalto-NMC) and the Cryohall (Low Temperature Laboratory) premises.


\end{document}